\input{aipcheck}

\documentclass[
    ,draft            
  ]
  {aipproc}

\layoutstyle{6x9}

\begin{document}

\title{Measuring Sparticles with the Matrix Element}


\author{Johan Alwall}{
  address={SLAC, Stanford University, Stanford, CA 94309, USA}
  ,altaddress={Department of Physics, National Taiwan University, Taipei, Taiwan} 
}

\author{Ayres~Freitas}{
  address={Department of Physics and Astronomy, University of Pittsburgh, PA 15260, USA}
}

\author{Olivier Mattelaer}{
  address={INFN, Sezione di Roma Tre,
and Universit\`a degli Studi Roma Tre, 00146, Roma, Italy }
  ,altaddress={Universit\'e Catholique de Louvain,
2, Chemin du Cyclotron, 1348 Louvain-la-Neuve, Belgium} 
}

\begin{abstract}
We apply the Matrix Element Method (MEM) to mass determination of squark
pair production with direct decay to quarks and LSP at the
LHC, showing that simultaneous mass determination of squarks and LSP
is possible. We furthermore propose methods for inclusion of QCD
radiation effects in the MEM.
\end{abstract}

\maketitle




The goal of the LHC at CERN, scheduled to start this year, is to
discover new physics through deviations from the Standard Model (SM)
predictions. After discovery of deviations from the SM, the next step
will be classification of the new physics.
An important first goal in this process will be establishing a mass
spectrum of the new particles. One of the most challenging scenarios
is pair-production of new particles which decay to invisible massive
particles, giving missing energy signals. Many methods have been
proposed for mass determination in such scenarios (for a recent list
of references, see e.g.~\cite{Matchev:2009iw}).



In this proceeding, we report the first steps in applying the Matrix
Element Method (MEM) in the context of supersymmetric scenarios giving
missing energy signals. 
After a quick
review of the MEM, we will focus on squark pair production, a
process where other mass determination techniques have difficulties to
simultaneously determine the LSP and squark masses. Finally, we will
introduce methods to extend the range of validity of the MEM,
by taking into account initial state radiation (ISR)
in the method.

The Matrix Element Method is a procedure to measure a set of
theoretical parameters from a sample of experimental events. It
associates to each event $x$ a weight $P( x | \alpha)$, which is
a measure of the probability for an event $x$ to be observed
given a set of parameters $\alpha$. The computation of
these weights is done by convoluting the
production of a parton-level configuration $y$, given by the squared
matrix element $|M_\alpha|^2( y)$, with the probability
that the parton-level configuration $y$ evolves into the reconstructed
event $x$, as modeled by a transfer function $W(x, y)$. As a
result, the weight of a specific event $x$ can be written as
\begin{equation}
\label{weight_def}
P(x | \alpha)=\frac{1}{\sigma_{\alpha}} \int d \phi(y) |M_{\alpha}|^2 (y) dq_1 dq_2 f_1(q_1) f_2(q_2) W(x, y)
\end{equation} 
where $f_1(q_1)$ and $f_2(q_2)$ represent the parton distribution functions, $d \phi(y)$ the phase space measure and $\sigma_{\alpha}$ the total cross section. The normalisation by the cross-section ensures that $P( x | \alpha)$ is a probability density: $ \int P( x |  \alpha) dx=1$.

In principle, this definition provides the best possible discriminator
on a event-by-event basis, and can be used to measure $\alpha$.
With a sample of $N$ events, we can construct a log-likelihood:
\begin{equation}
\label{likelihood_3}
-\ln (L)=-\sum_{i=1}^{N} \ln(P(x_i|\alpha)) + N \int \mathrm{Acc}(x) P(x|\alpha)dx
\end{equation}
where the acceptance term $\mathrm{Acc}(x)$, corrects for the bias introduced
by detector acceptance and event selection.
The set of parameters $\alpha$ maximizing the likelihood correspond to
the most probable value.


Although the MEM is conceptually very simple, the
numerical evaluation of the weights is not straightforward due to the
large variations of the integrand. Indeed, both the square
matrix-element $|M_{ \alpha}|^2$ and the transfer function $W( x,
y)$ are highly non uniform on the phase-space. In consequence, a
specific Monte Carlo phase-space integrator, designed to deal with the
specific behavior of the integrand is required.

MadWeight \cite{MW} is a publicly available phase space generator
dedicated to perform this type of integration. In order to find the
best phase-space mapping, MadWeight applies a series of local changes
of variables to promote, when possible, the invariant mass of resonant
propagators as variables of integrations. In this way, MadWeight is
able to find the best phase-space mapping for any decay chain and a
large class of transfer functions.


\section{Squark Measurment}

The MEM has been successfully applied for determination of the top
quark mass at the Tevatron \cite{Abulencia:2006mi}. However, applying
the MEM to 
supersymmetric scenarios brings new difficulties, including a large
parameter space, a priori unknown order of particles in decay chains,
and unknown masses of invisible particles. In the preliminary study
presented here, we focus on the latter of these difficulties by
studying squark pair production, where the squarks decay directly to
quark and LSP, the lightest neutralino. The signature is two hard jets
and large missing transverse energy. This scenario is interesting,
because the $m_{T2}$ ``kink'' method
\cite{Barr:2009jv}, as well as other methods for
simultaneous reconstruction of the squark and neutralino masses
often fails for two-body decays of the squarks.

\begin{figure}
  \includegraphics[width=0.33\textwidth]{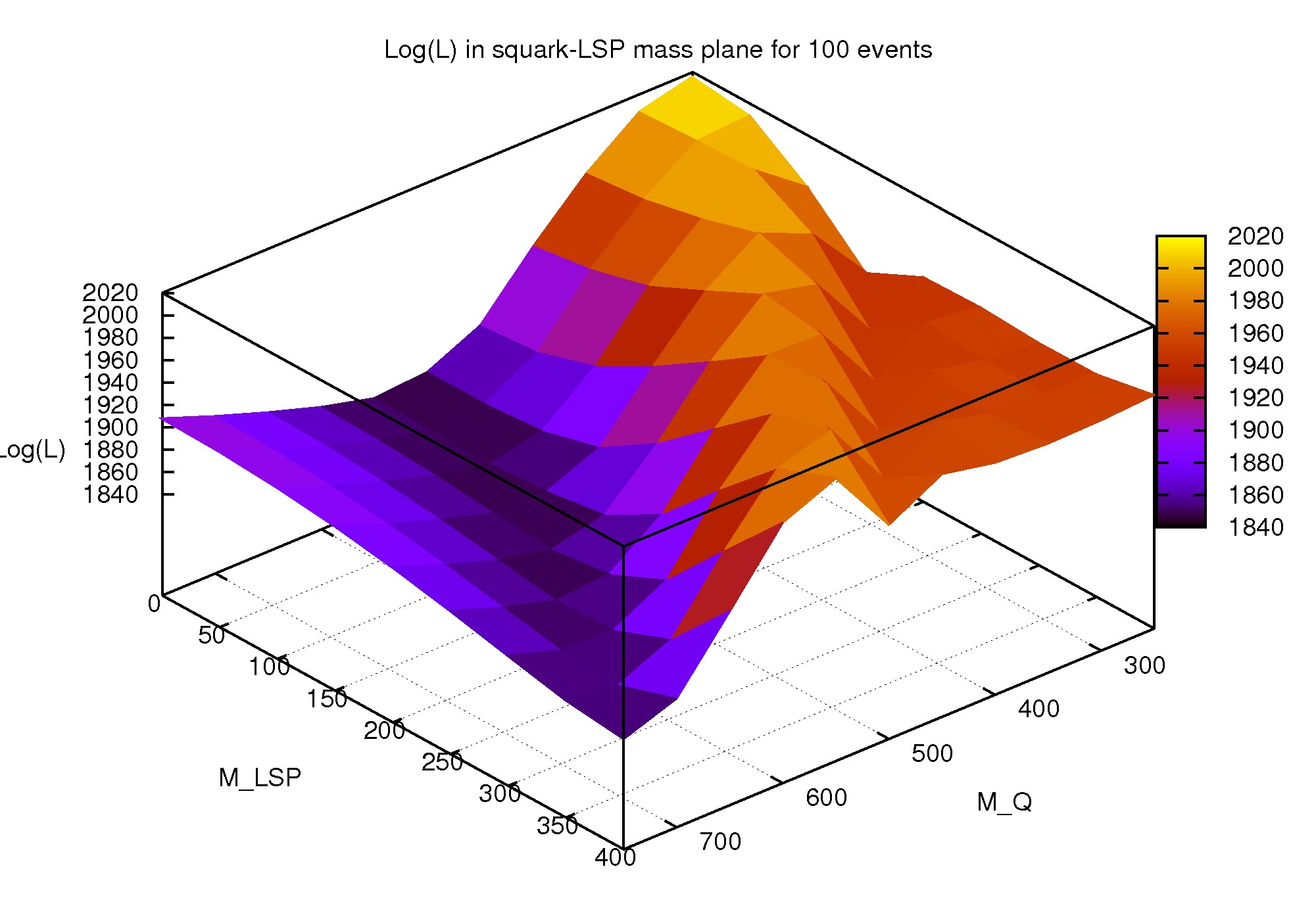}
  \includegraphics[width=0.33\textwidth]{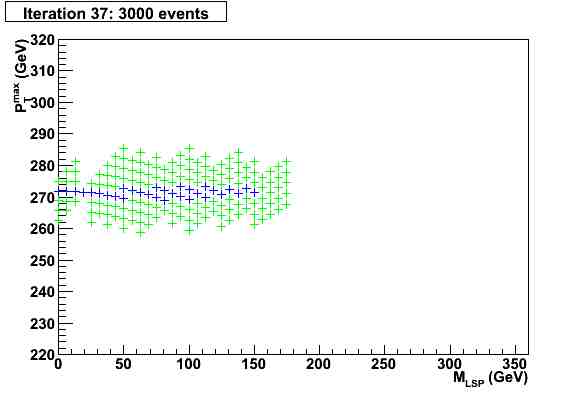}
  \includegraphics[width=0.33\textwidth]{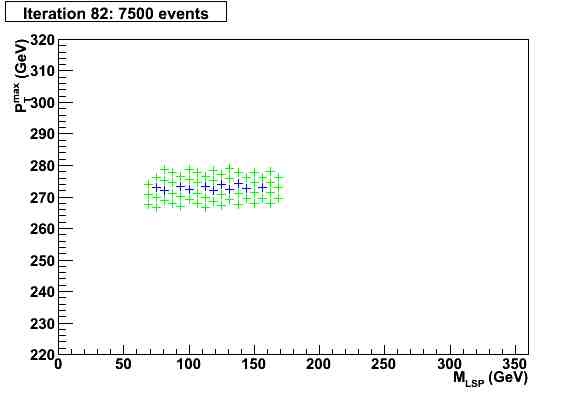}
  \caption{\label{fig:sqsq-nocuts} Matrix element $-\log(L)$ for a) 100, b)
3000 and c) 7500 parton level $\tilde q\bar{\tilde q}\to q \bar q
\tilde\chi^0\tilde\chi^0$ events at 
the LHC. 
The true values are at
$(m_{\tilde q},m_{\tilde\chi^0})=(561,97)$ GeV. b) and c) show the
valley in the $(m_{\tilde\chi^0},p_T^\mathrm{max})$ plane. 
The blue
crosses show parameter points with $-\log(L)$ at most 4 above $-\log(L)_\mathrm{min}$
(corresponding to approximately 90\% CL). 
}
\end{figure}

As a first check that the Matrix Element method can indeed reconstruct
masses in this scenario, we simulate squark pair production at the
LHC, and study this process at parton level without cuts, with
$m_{\tilde q}= 561$ GeV and $m_{\tilde\chi^0} = 97$ GeV. The resulting
negative logarithmic likelihood function is shown for 100 events in
Fig.~\ref{fig:sqsq-nocuts}a). The minimum of $-\ln(L)$ forms a
``valley'' in the $(m_{\tilde q},m_{\tilde\chi^0})$ plane. This valley
is closely aligned with the expression for the energy of the quark in
the rest frame of the decaying squark, i.e., the maximum $p_T$ of the
quark produced by squark production at rest, given by
$
p_T^\mathrm{max}=(m_{\tilde q}^2-m_{\tilde\chi^0}^2)/2m_{\tilde q}
$.
Fig.~\ref{fig:sqsq-nocuts} b) and c) show the extent of the valley
using 3000 and 7500 events, respectively (corresponding to an
integrated luminocity of 10 fb${}^{-1}$ and 25 fb${}^{-1}$
respectively, assuming 100\% branching ratio). With 3000 events, only
an upper limit on $m_{\tilde\chi^0}$ can be determined. For 7500
events, however, the mass range for the $\tilde\chi^0$ is already
quite restricted, and is determined to be between 65 GeV and 160
GeV. The $p_T^\mathrm{max}$ value is very well determined, with an
uncertainty of $\pm 2$ GeV (less than 1\%).

Further studies indicate that these results are still valid
when realistic cuts, and the effect
of hadronization and detector simulation, are considered, although the
efficiency is then reduced. 
See \cite{paper}.

\section{Initial state radiation}

One challenge associated with the Matrix Element Method is 
how to deal with initial state radiation (ISR). Energetic
jets radiated from the intial state partons will be much more
ubiquitous for high-$p_t$ physics at the LHC than at the Tevatron
\cite{Plehn:2005cq}. To simply ignore their effect leads to bad or
impossible fits of the matrix element algorithm; on the other hand, a
very strict veto on additional jets (besides jets expected from the
primary hard process) would cause a significant reduction of signal
statistics.

Conceptually, the most straightforward way of including events with sizeable ISR
would be the use of matrix elements  with additional partons in the final state.
In practice, however, there is a limit to the multiplicity of
the matrix elements due to the computing time for the matrix element generation
and the phase-space integration for the weight computation. Therefore, in most
cases, only one or two jets from radiation can be included. 

\begin{figure}
  \includegraphics[width=0.5\textwidth]{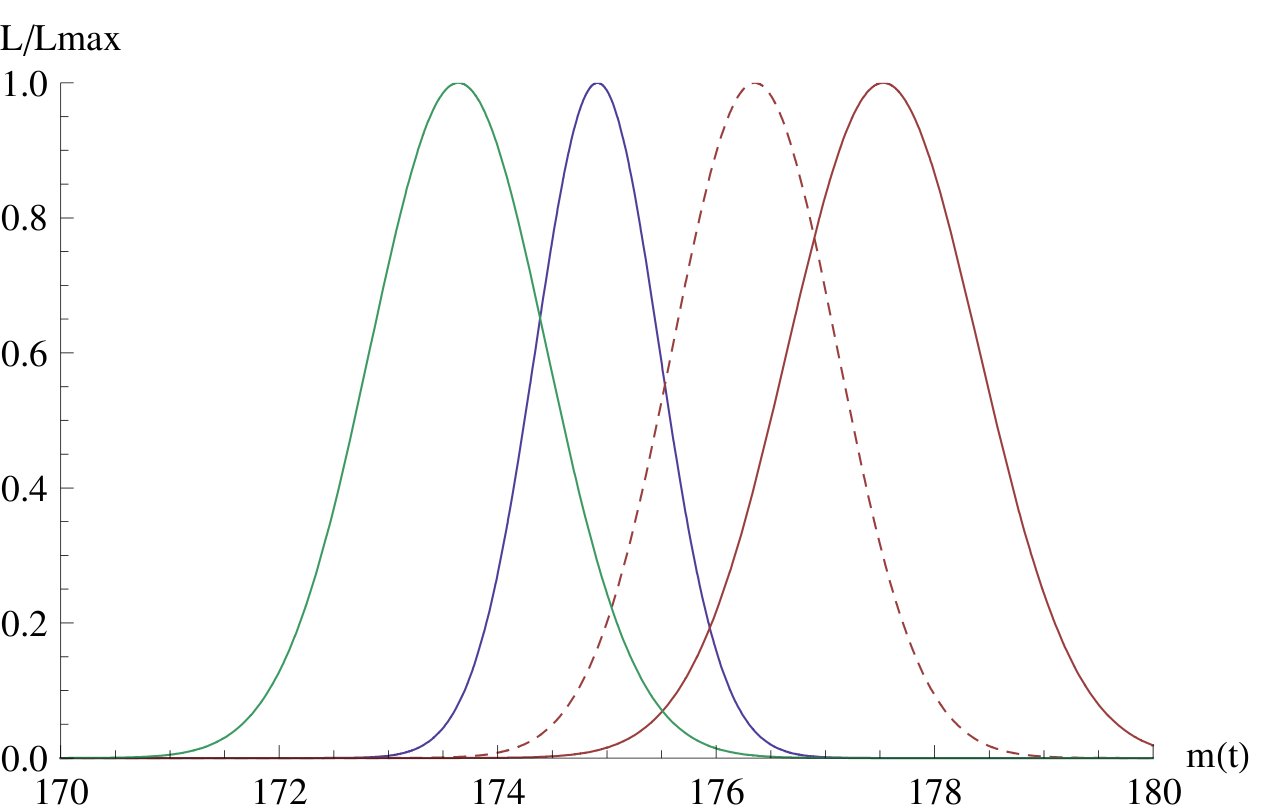}
  \caption{\label{fig:isr} Reconstruction of the top quark mass from a matrix
  element likelihood fit to 1000 di-lepton top-antitop events at the LHC with
  $\sqrt{s} = 14$~TeV. A top mass of $m_t = 175$~GeV has been used for the event
  generation. {\sl Blue:} ``ideal'' case
  without ISR in the event generation. {\sl Red solid (rightmost
  curve):} realistic event 
  generation with ISR, but events with extra jets with $p_T > 40$~GeV
  have been vetoed. 
  {\sl Red dashed:} same as above but also including $t\bar{t}$ matrix elements
   with one additional jet. {\sl Green:} boost method to deal with ISR in the
   events.} 
\end{figure}

This situation is illustrated in Fig.~\ref{fig:isr} for the example of
1000 simulated events of
di-lepton top pair production at the LHC, $pp \to t\bar{t} + nj \to
b\bar{b}l^+l^-\nu_l\bar{\nu}_l + nj$, where $nj$ indicates $n$ jets from
ISR. The events have been generated with Pythia 6.4/PGS4 \cite{Sjostrand:2006za,PGS4}.
Backgrounds have not been included.
As is evident from the figure, the inclusion of matrix elements with one extra jet
already significantly improves the agreement (peak of likelihood curve) with the
true input value $m_t = 175$~GeV and also the
statistical precision (width of likelihood curve). 

We also propose a different method to treat ISR. We observe that the
most significant effect of ISR is on the kinematics of the events,
since without proper inclusion of ISR the momentum balance would be
violated. This effect can be taken into account by simply boosting the
hard event by the momenta of the initial state jets.
The longitudinal incoming momenta are integrated over in the computation
of the likelihood (see \eqref{weight_def}), so
it is sufficient to perform the boost for
the transverse coordinates only.
Since the identification of the $b$ jets from top decay is not unique,
we sum over all permutations of the four hardest jets in the event.

The result of a fit with this boost method is shown by the green curve in
Fig.~\ref{fig:isr}. It agrees with the true input value within statistical
errors. The quality of the fit result is comparable to the computation with
explicit $t\bar{t}+j$ matrix element (red-dashed curve), but requires less
computation time. Furthermore, the result of the boost method can be improved
by including QCD splitting functions Sudakov form factors for the (cumulative)
ISR on each leg \cite{paper}.


In summary, we have here presented a first study in applying the
Matrix Element Method to supersymmetric particle production with decay
to massive invisible particles. In particular, we showed that it is
possible to simultaneously determine the squark and LSP masses in
squark pair production with exclusive decay to quark and LSP. We also
proposed two methods to include the effects of initial state radiation
in MEM calculations. For more details we refer to \cite{paper}.


{\bf Acknowledgements:} This research was supported in part by Computational Resources on PittGrid
(\texttt{www.pittgrid.pitt.edu}).


\bibliographystyle{aipproc}   

\bibliography{matrixelements}

\IfFileExists{\jobname.bbl}{}
 {\typeout{}
  \typeout{******************************************}
  \typeout{** Please run "bibtex \jobname" to optain}
  \typeout{** the bibliography and then re-run LaTeX}
  \typeout{** twice to fix the references!}
  \typeout{******************************************}
  \typeout{}
 }






\end{document}